\documentclass[review,endfloat]{elsarticle}

 \usepackage{graphics}
 \usepackage{graphicx}
\usepackage{textgreek}
\usepackage{subcaption} 
\journal{Applied Radiation and Isotopes}

\begin{document}

\begin{frontmatter}

\title{Measurement of $^{227}\mathrm{Ac}$ Impurity in $^{225}\mathrm{Ac}$ using Decay Energy Spectroscopy}

\author[label1]{A. D. Tollefson}
\address[label1]{Los Alamos National Laboratory, Los Alamos, NM 87545, USA}
\address[label2]{NIST Boulder Laboratories, Boulder, CO 80305, USA}
\address[label3]{University of Colorado, Boulder, CO 80309, USA}



\author[label1]{C. M. Smith\footnote{\textit{Corresponding author}, email: \textbf{cmsmith@lanl.gov}}}
\author[label1]{M.~H.~Carpenter}

\author[label1]{M.~P.~Croce}

\author[label1]{M.~E.~Fassbender}

\author[label1]{K.~E.~Koehler}

\author[label1]{L.~M.~Lilley}

\author[label1]{E.~M.~O'Brien}

\author[label2]{D.~R.~Schmidt}

\author[label1]{B.~W.~Stein}

\author[label2,label3]{J.~N.~Ullom}

\author[label1]{M.~D.~Yoho}

\author[label1]{D.~J.~Mercer}

\begin{abstract}
$^{225}\mathrm{Ac}$ is a valuable medical radionuclide for targeted $\alpha$ therapy, but $^{227}\mathrm{Ac}$
 is an undesirable byproduct of an accelerator-based synthesis method under investigation. Sufficient detector sensitivity is critical for quantifying the trace impurity of $^{227}\mathrm{Ac}$, with the $^{227}\mathrm{Ac}$/$^{225}\mathrm{Ac}$ activity ratio predicted to be approximately 0.15\% by end-of-bombardment (EOB). Superconducting transition edge sensor (TES) microcalorimeters offer high resolution energy spectroscopy using the normal-to-superconducting phase transition to measure small changes in temperature. By embedding $^{225}\mathrm{Ac}$ production samples in a gold foil thermally coupled to a TES microcalorimeter we can measure the decay energies of the radionuclides embedded with high resolution and efficiency. This technique, known as decay energy spectroscopy (DES), collapses several peaks from $\alpha$ decays into single Q-value peaks. In practice there are more complex factors in the interpretation of data using DES, which we will discuss herein. Using this technique we measured the EOB $^{227}\mathrm{Ac}$ impurity to be (0.142 $\pm$ 0.005)\% for a single production sample. This demonstration has shown that DES can distinguish closely related isotopic features and is a useful tool for quantitative measures.
 \end{abstract}


\end{frontmatter}


\section{Introduction}

\subsection{$^{\textit{225}}\mathrm{\textit{Ac}}$ Based Radiopharmaceuticals}

Radionuclides are powerful tools for oncological treatments that were first introduced to market in the early 2000s \citep{Thiele2018}. Early radionuclide treatments relied on $\beta^{-}$ emitters, and only within the past decade have alpha emitters entered the marketplace for potential cancer treatments despite greater cytotoxicity. Choosing an appropriate radionuclide for treatment depends on its half-life, decay energy, type of decay, and chemistry, as all these factors contribute to its effectiveness and practicality. Additionally, availability and production cost are pragmatic factors that must be considered. 

\begin{figure}[!h]
  \begin{subfigure}[b]{0.45\textwidth}
  \centering
    \includegraphics[width = 3.8cm]{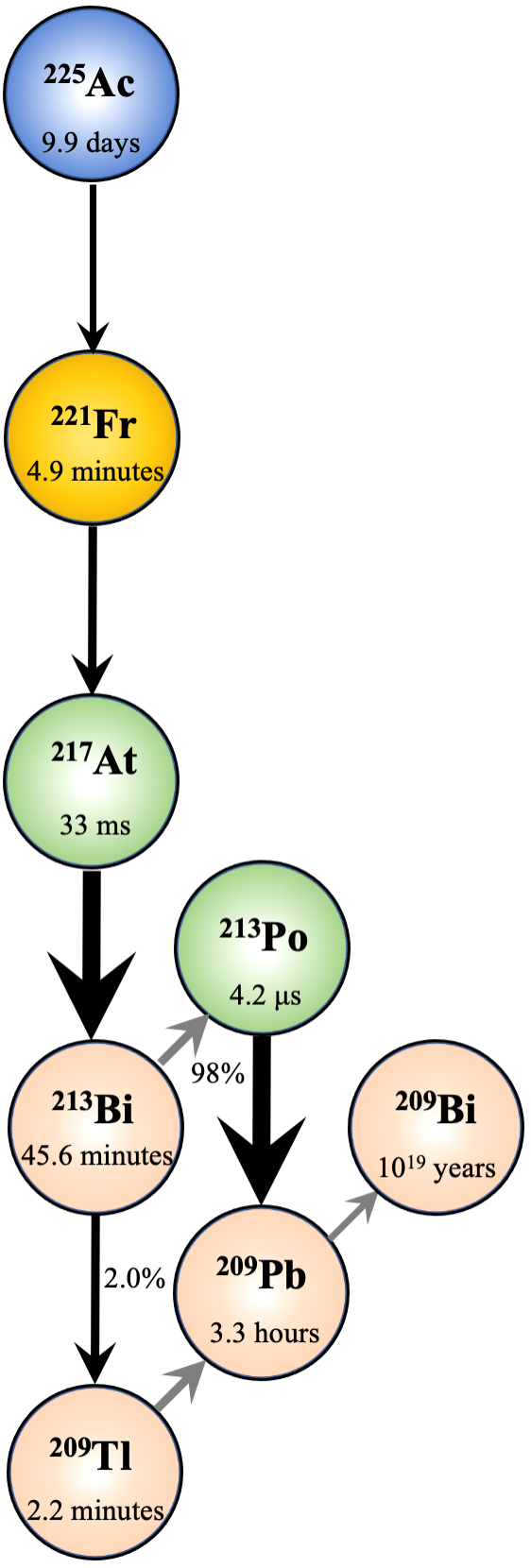}
    \caption{\label{fig:225DecayChain} $^{225}\mathrm{Ac}$ decay chain. This decay chain indicates the emission of four $\alpha$ particles in rapid succession.}
  \end{subfigure}
  \hspace{5mm}
  \begin{subfigure}[b]{0.45\textwidth}
  \centering
    \includegraphics[width = 5.2cm]{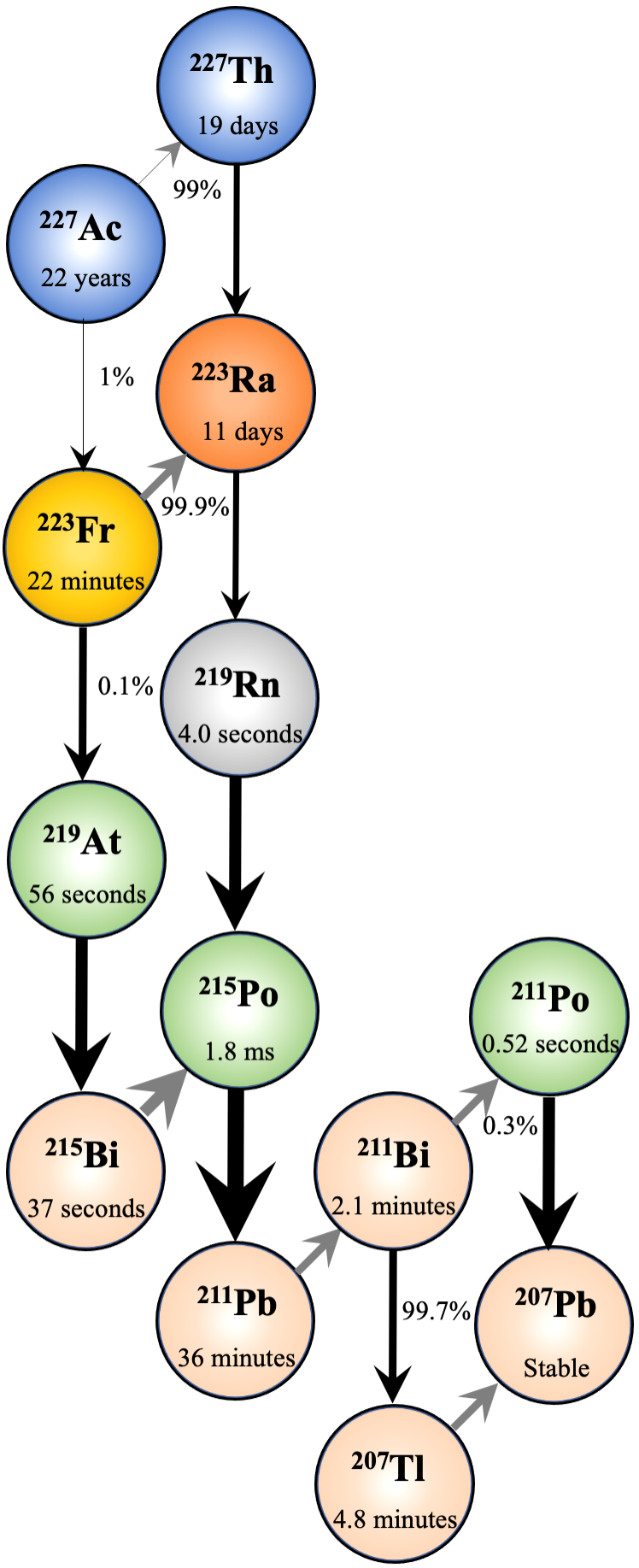}
    \caption{\label{fig:227DecayChain} $^{227}\mathrm{Ac}$ decay chain. This complex decay chain contributes to the complex decay energy spectrum.}
  \end{subfigure}
  \caption{\label{DecayChains} Decay chains of $^{225}\mathrm{Ac}$ and $^{227}\mathrm{Ac}$. Gray arrows indicate $\beta^{-}$ decay to the next nuclide, black arrows indicate $\alpha$ decay. The thickness of the arrow indicates the parent half-life, with thicker arrows corresponding to shorter half-lives. }
\end{figure}

One of the more promising alpha emitters currently under investigation is $^{225}\mathrm{Ac}$. Patients that have undergone $^{225}\mathrm{Ac}$ treatment trials have exhibited significant cancer remission \citep{Makvandi}. A property of $^{225}\mathrm{Ac}$ that makes it an interesting candidate is the rapid sequential emission of four alpha particles (Figure \ref{DecayChains}), each depositing a large quantity of energy (greater than 5 MeV) over a short distance, and causing critical double-strand breaks in DNA \citep{Birnbaum,Makvandi}. Double-strand DNA breaks are irreparable and ultimately lead to cell death. $\alpha$ radiation causes more damage to individual cancer cells than $\beta^{-}$ radiation, to which cells can potentially acquire resistance \citep{Makvandi}. $^{225}\mathrm{Ac}$ has a 9.92 day half-life, which is enough time for processing and distribution as well as time to bind the radioisotope to biological targeting vectors, such as antibodies or peptides, which gives extended in vivo circulation time for targeted therapy\citep{Thiele2018}. Binding radioisotopes to targeting vectors increases the effectiveness and safety of this treatment by selectively target cells for destruction \citep{Makvandi,McDevitt1998}. This radionuclide also has potential antibiotic and antifungal applications \citep{Dadachova2010,Nosanchuk2012}.


\pagebreak

\subsection{$^{\textit{227}}\mathrm{\textit{Ac}}$ Impurity}

Although $^{225}\mathrm{Ac}$ is ideal for targeted $\alpha$ therapy, it is not naturally occurring, so its scarcity prevents its widespread use in clinical therapy. The annual US production, which is based on $^{229}\mathrm{Th}$ extracted from legacy $^{233}\mathrm{U}$ stockpiles, only provides around 1.7 Ci worth of material which is not enough to meet global demand \citep{Thiele2018}. The Tri-Lab Effort between Los Alamos National Lab, Brookhaven National Lab, and Oak Ridge National Lab is studying accelerator-based synthesis to increase supply \citep{Morgenstern2012} using an $^{232}\mathrm{Th}$(p,x)$^{225}\mathrm{Ac}$ reaction. However, a trace radioactive byproduct of accelerator synthesis that cannot be chemically separated is $^{227}\mathrm{Ac}$, which has a half-life of 22 years. $^{227}\mathrm{Ac}$ is undesirable since it outlives its targeting vector binding time and tends to accumulate in the liver, where it may be toxic \citep{taylor1970metabolism}. Due to potential challenges associated with facility licensing and decommissioning, waste disposal, and product licensing, low concentrations of $^{227}\mathrm{Ac}$ must be confirmed in bulk $^{225}\mathrm{Ac}$ samples. 

Conventional $\alpha$-spectroscopy will struggle with detection of $^{227}\mathrm{Ac}$ shortly after the sample is removed from the beam due to an unresolved forest of $\alpha$ particle energies, low relative signal from the $^{227}\mathrm{Ac}$ chain, as well as $^{219}\mathrm{Rn}$ escape from the $^{227}\mathrm{Ac}$ decay chain. $\gamma$ spectroscopy may be applicable as a quantification method, but is likewise limited by the complex spectrum and low intensity of the $^{227}\mathrm{Ac}$ chain signatures at early times. This paper describes the application of decay energy spectroscopy (DES), a novel radiometric technique, to determine the isotopic composition of a chemically-purified $^{225}\mathrm{Ac}$ production sample. This technique offers a cleaner spectrum, high sensitivity, and requires less than 1 Bq of material.

\section{Experimental Setup}

DES is a novel approach that overcomes many challenges present in current analytic methods. The underlying physical principle is simple; an ultra-sensitive microcalorimeter measures and quantifies the energy from every discrete decay event in a sample. In this manner, DES has the potential to provide both high resolving power and absolute activity in a single measurement \citep{Hoover2015,Lee2010}.
 
\begin{figure}[hbt]
 
\centering
\includegraphics[width = \linewidth]{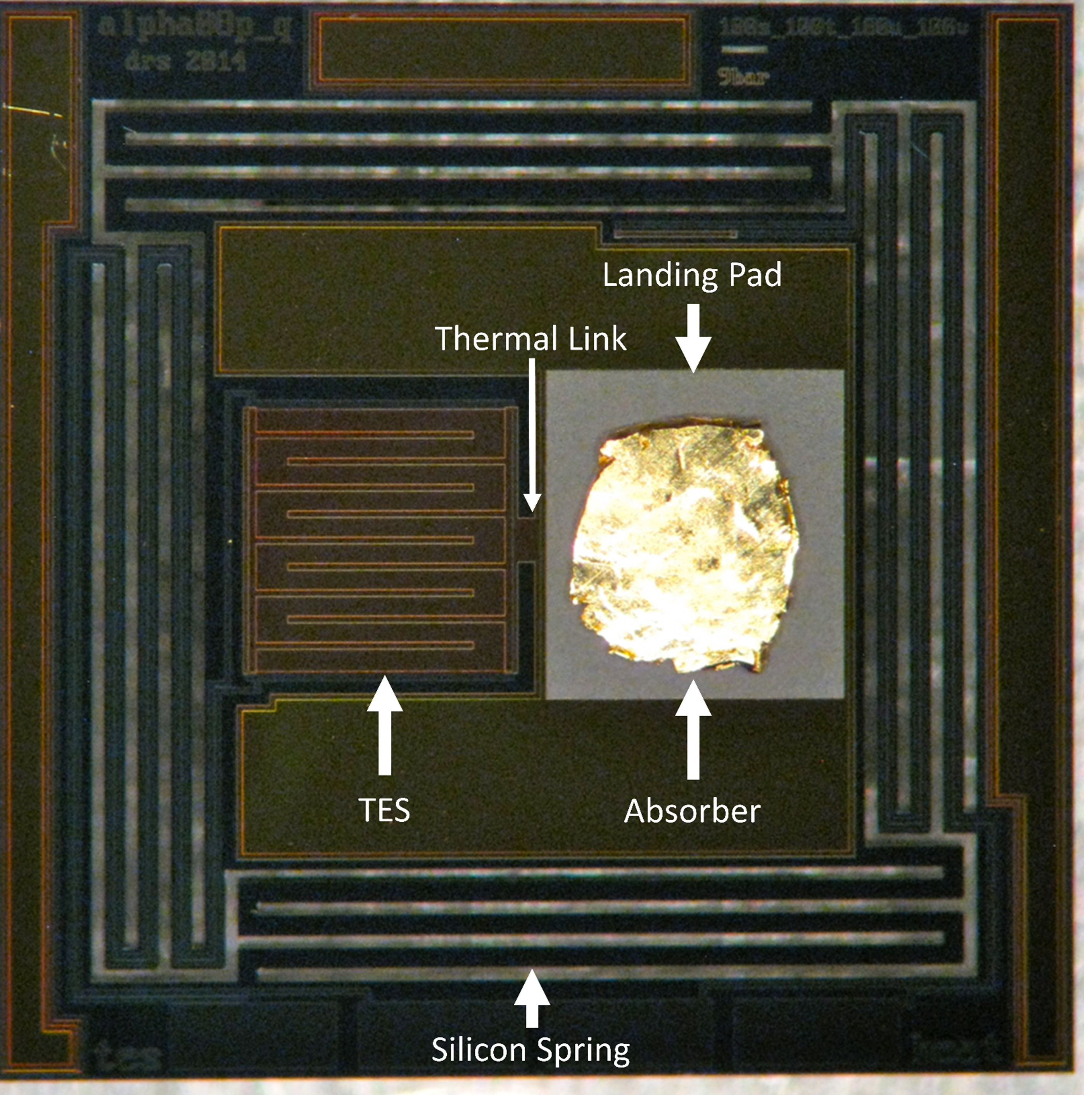}
\caption{\label{fig:TES_photo} A labeled TES (left) with gold absorber (right). The sample is embedded in the absorber to achieve 100\% detection efficiency for $\alpha$ decays. The silicon spring is present to alleviate mechanical strain when handling.}
\end{figure}

Microcalorimeters for DES are operated at ultra-low temperatures to reduce the calorimeter component heat capacity and background thermal noise. The devices consist of three main parts: a calorimeter platform, a sensitive thermometer, and an absorber with embedded radionuclides. In this work, the calorimeter platform is formed from micro-machined silicon and the platform is directly integrated with a superconducting transition edge sensor (TES) thermometer. The absorber is formed from a thin gold foil, with radionuclides deposited from solution on the foil. The foil is then folded to encapsulate the radionuclide and mechanically alloyed to break up crystals from the deposition and to uniformly distribute the radionuclides throughout the whole of the foil \citep{Hoover2015}. The absorber with embedded radionuclides is attached to a gold landing pad via an indium bump bond, which is electroplated to a silicone wafer. An example of the detector chip is seen in Figure~\ref{fig:TES_photo}.

The detector chips are operated inside an adiabatic demagnetization refrigerator (ADR). The TES used for these measurements is a Mo/Cu bilayer with a superconducting critical temperature of around 100 mK and the ADR temperature is controlled at 80 mK.  The TES is electrically biased and self-heats into the superconducting transition.  Decay events deposit energy into the absorber and the energy dissipates from the absorber through the thermal link to the TES. The rise in temperature of the TES causes the resistance of the TES to change and the resulting current change signal is amplified using an initial superconducting quantum interference device (SQUID). This signal is then further amplified by a series SQUID array and a room-temperature amplifier. The event signals are then digitized and archived for analysis. The TES is weakly thermally coupled to the surrounding low-temperature bath, with a return to thermal equilibrium occuring around 200 ms after a decay event. Data runs lasted between 12 and 30 hours, limited by cryostat hold time. The system houses eight separate channels with each detector operating independently \citep{Croce2009}.

\section{Sample Preparation}

Spectra were obtained from four sample types: 1) a pure sample of $^{225}\mathrm{Ac}$ from a $^{229}\mathrm{Th}$ generator; 2) a pure sample of $^{227}\mathrm{Ac}$; 3) synthetic mixtures of the above to simulate 1\% and 10\% $^{227}\mathrm{Ac}$/$^{225}\mathrm{Ac}$ activity ratios; and 4) an $^{225}\mathrm{Ac}$ production sample, which is from a $^{232}\mathrm{Th}$ target irradiated at LANL with an 100 MeV proton beam for 46 hours and chemically purified at ORNL \citep{Robinson2020}. ORNL also sourced the $^{229}\mathrm{Th}$ and $^{227}\mathrm{Ac}$ for samples 1-3. Samples 1-3 were measured to better understand what spectral features might appear in the more valuable production sample prior to its measurement. Only the production sample is reported on in this publication.     
   
Low activity ($^{229}\mathrm{Th}$: 0.87 Bq/$\mu$L, $^{227}\mathrm{Ac}$: 0.38 Bq/$\mu$L) working stocks for microcalorimetry measurements were leached (8M HNO$_{3}$, Fisher Optima Grade prepared in Teflon distilled 18 M$\Omega$ H$_{2}$O) from glass vials from previous experimental campaigns. Radionuclides were purified using previously published methods\citep{Ferrier2016,Ferrier2017,Ferrier2018}. Stock material was characterized with a high-purity germanium detector measurements. The production sample was found to have approximately a 1:9 activity ratio of $^{227}\mathrm{Ac}$ to $^{225}\mathrm{Ac}$ when measured at 40 days since end-of-bombardment (EOB). Quantification by $\gamma$ rays, however, is challenging for high-purity production samples measured shortly after irradiation and chemical processing because the background from $^{225}\mathrm{Ac}$ progeny is high, and the characteristic signatures of the $^{227}\mathrm{Ac}$ chain take a few weeks to develop.  Samples were prepared by pipetting 0.5-2.0 $\mu$L of solution onto 5 $\mu$m-thick Au foils (99.9\% purity Au by mole), followed by five minutes of drying with a heat lamp. The Au foils were then mechanically alloyed with pliers to distribute the embedded radionuclide and break up any crystalline deposits. This process has been found to improve resolution \citep{Hoover2015}.

\section{Method}

\subsection{Data Processing}
\begin{figure}[h!]
\centering
\includegraphics[width = \textwidth]{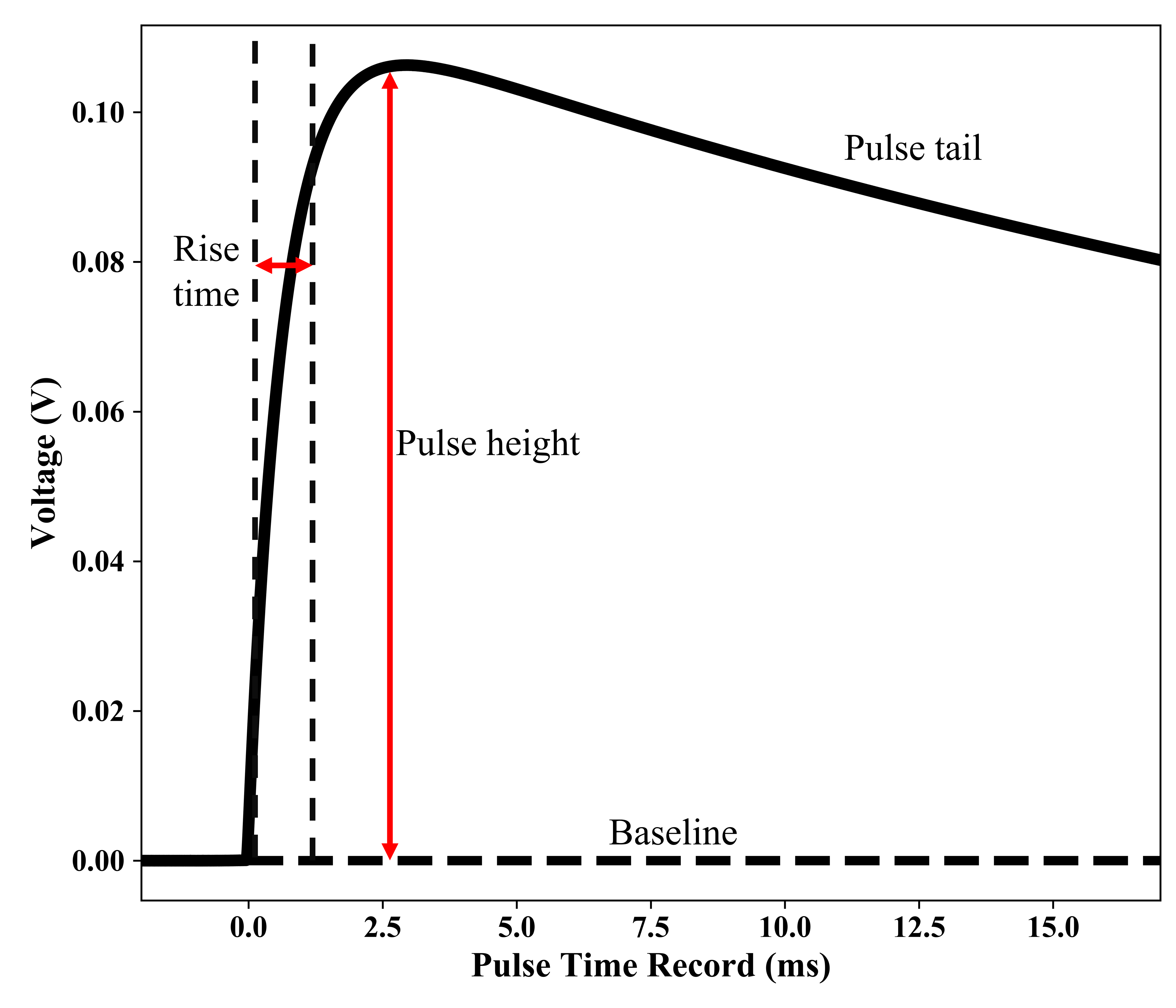}
\caption{\label{fig:Pulse} An example of a normal pulse, with important characteristics labeled. The pulse rise time is defined as the time between 10\% of the pulse height and 90\%, the baseline is the steady-state pre-trigger voltage, and the tail is the voltage signal as the TES returns to thermal equilibrium with the surrounding bath. The tail is characteristic of the detector decay time.}
\end{figure}

Triggered pulse records of 20 ms length are recorded from the TES signal stream. The sample rate is 204800 samples per second for 4096 samples per pulse record. These pulse records are preserved in entirety and written to disk. An example of a normal pulse can be seen in Figure \ref{fig:Pulse}.

\begin{figure}[h!]
\centering
\includegraphics[width=\textwidth]{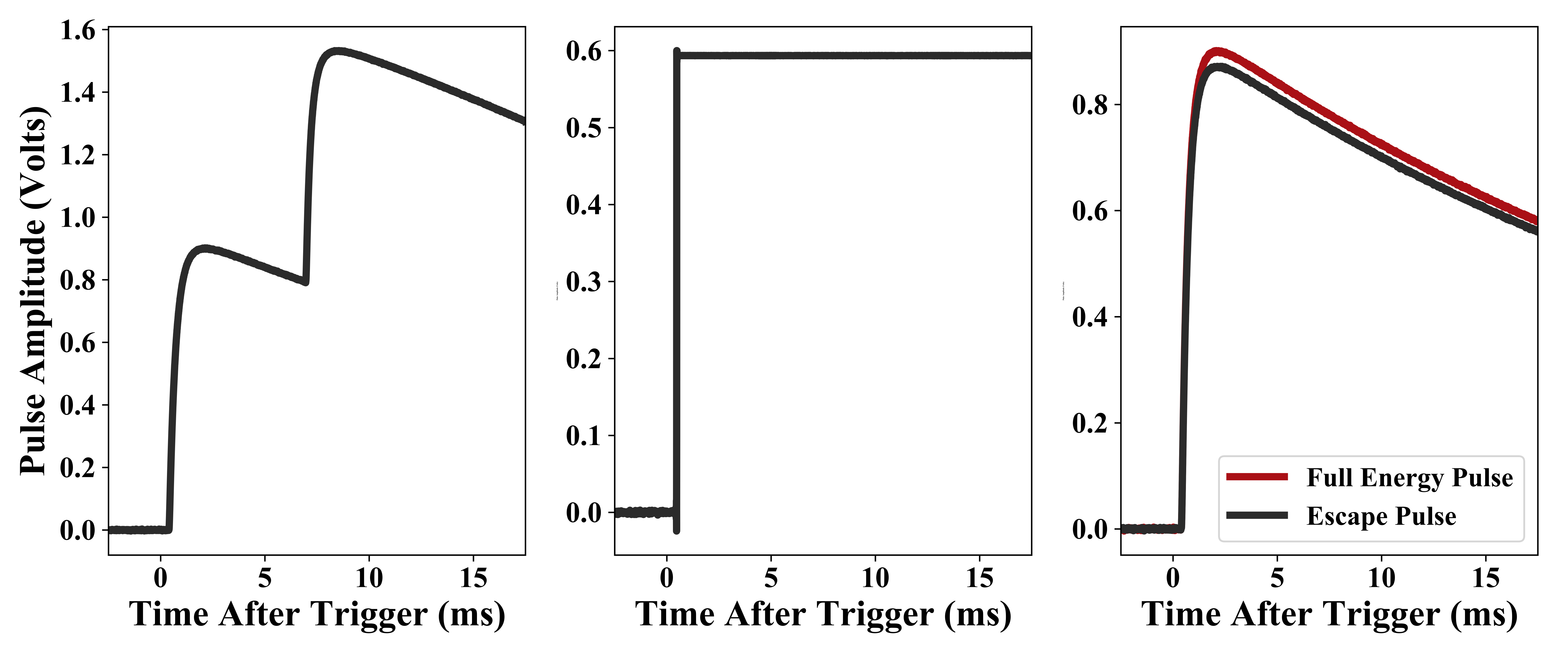}
\caption{\label{fig:AnomalousPulses} An example of various anomalous pulses produced in DES. (Left) Pulse pileup. The second pulse arriving during the record of the first pulse. These lead to high energy, poorly resolved spectral features as well as high-energy background. (Center) Flux jump. These occur in the SQUID used for TES readout and contribute to low-energy background. (Right) Full energy and Escape Pulses. The full energy pulse (red) is from the decay of $^{211}\mathrm{Bi}$, and the escape pulse (black) is from another $^{211}\mathrm{Bi}$ decay with a 351 keV photon escaping. These small changes in pulse height lead to multiple peaks corresponding to the same decay, complicating the spectra.}
\end{figure}

 In DES, there can be a significant number of pulses not directly proportional to the Q-values of corresponding decay events. This is due to low-probability $\gamma$ or x-ray escapes, non-physical events such as flux jumps characteristic of SQUID amplifier operations, and pulse pileup (Figure \ref{fig:AnomalousPulses}). Escape events complicate the spectrum, but are identifiable by their low amplitude and distance from the corresponding full energy peak and can be quantified. Flux jumps have a distinguishable pulse shape and are easily removed from the dataset. Pulse pileup refers to two decay events happening faster than the decay time of a given pulse and yields high-energy, poorly resolved peaks. These were mitigated by applying cuts to pulse characteristics. In particular, pulses with anomalously low rise times, where rise time is defined as the time between 10\% and 90\% of a pulse's maximum height, were indicative of flux jumps, and pulses with long rise times were pileup. Pileup was originally rejected outright, but the presence of valuable pileup features for quantification led us to maintain these pulse records. Optimal Wiener filtering was then applied to the surviving pulse records, and the filtered values were drift-corrected using temporal shifts to compensate for temperature fluctuations and shifts in pre-trigger pulse baseline \citep{1643650}.

\subsection{Peak Identification} 
  
\begin{figure}[h!]
\centering
\includegraphics[width = \linewidth]{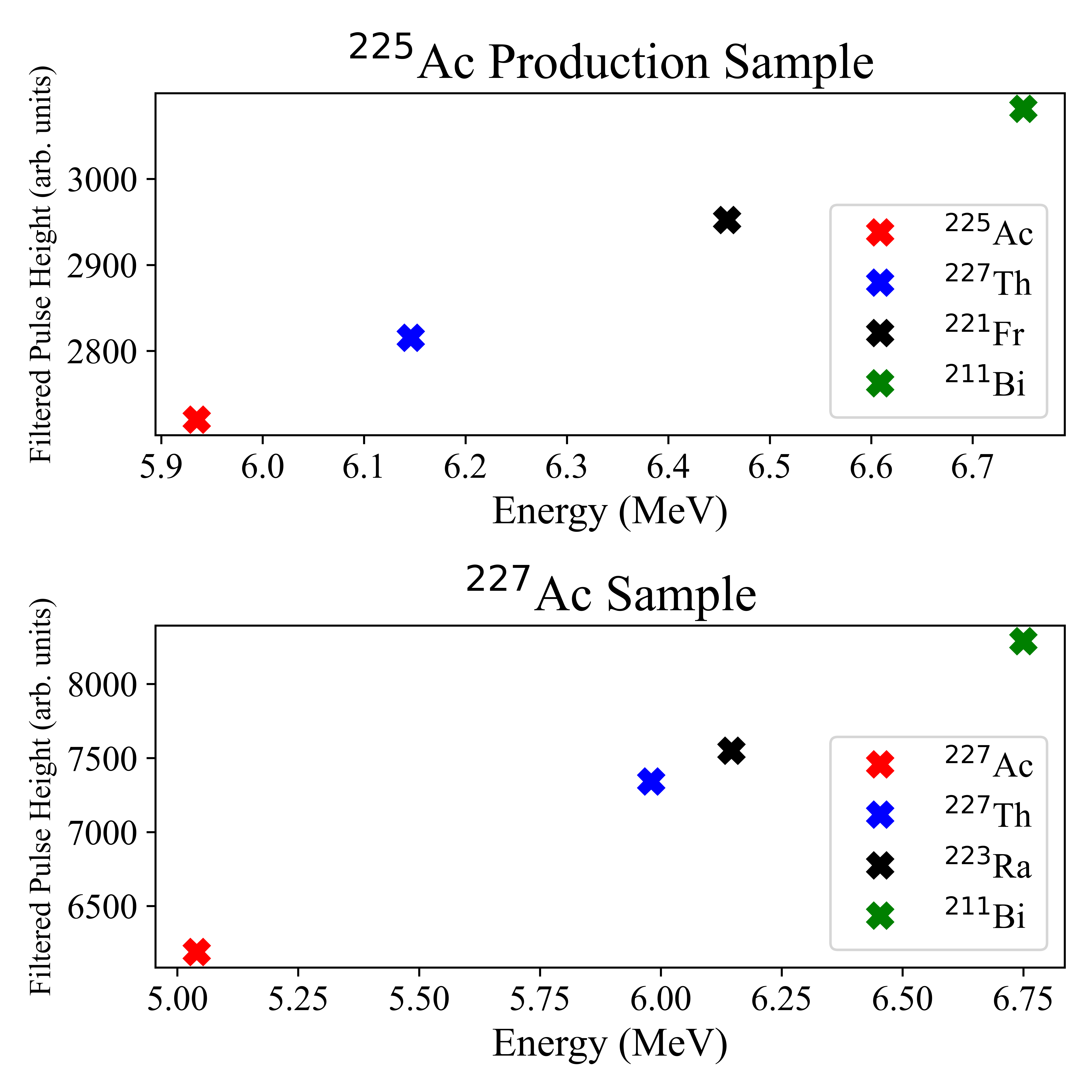}
\caption{\label{fig:Calibration} Plot of the filtered pulse height value of the peak centroids versus the energy of the decay. These plots indicate the fairly linear response of the detector following a spline fit in the direct Q value energy range. This allowed us to extrapolate energy calibration to the high-energy cascade region, aiding identification of these features.}
\end{figure}

\begin{figure}[h!]
\centering
\includegraphics[width = \linewidth]{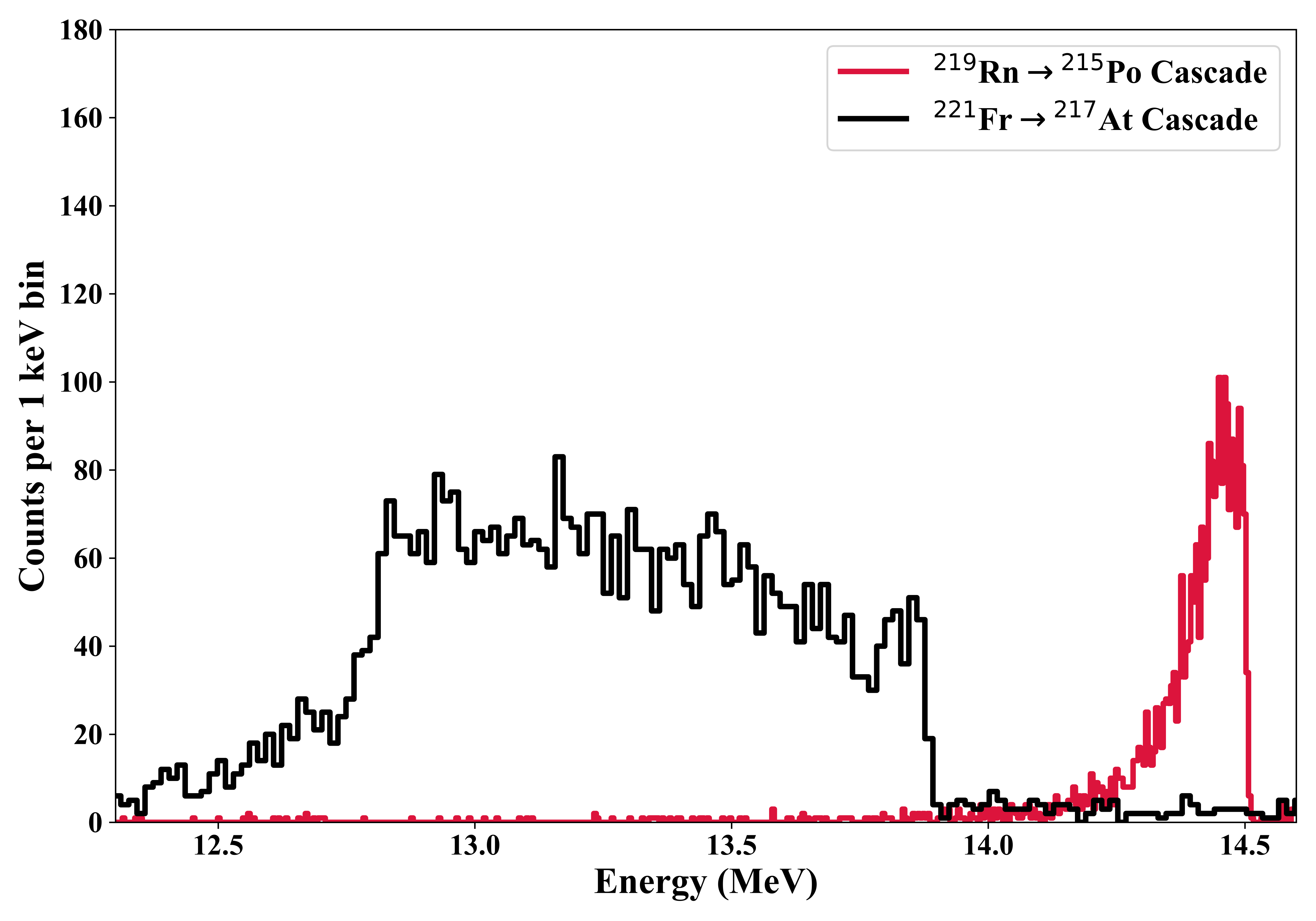}
\caption{\label{fig:Cascades} Plot of high energy cascades present in overlaid $^{225}\mathrm{Ac}$ and $^{227}\mathrm{Ac}$ spectra. The $^{221}\mathrm{Fr}\rightarrow{}^{217}\mathrm{At}$ cascade is present in the $^{225}\mathrm{Ac}$ spectra and the $^{219}\mathrm{Rn}\rightarrow{}^{215}\mathrm{Po}$ cascade is present in the $^{227}\mathrm{Ac}$ spectrum. All data can be found in the citation \citep{Nudat}.}
\end{figure}

\begin{figure}[hbt]
\centering
\includegraphics[width=\linewidth]{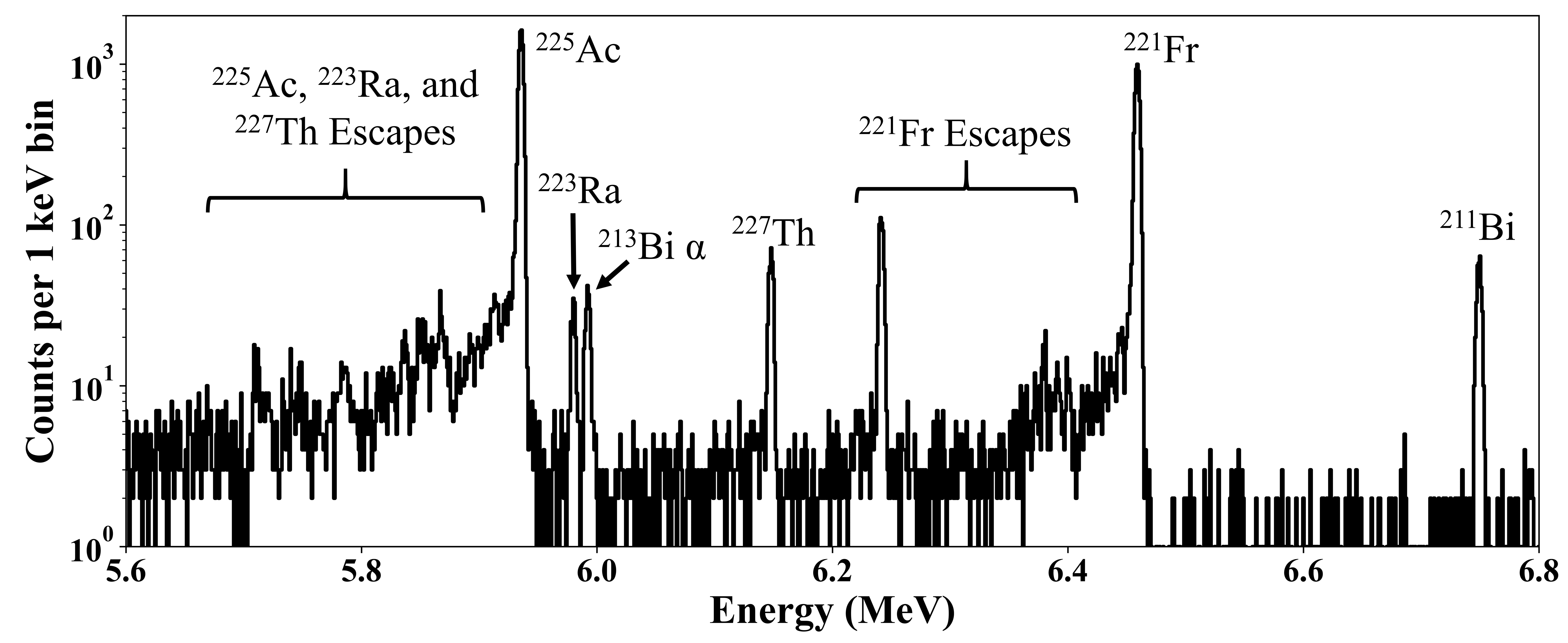}
\caption{\label{fig:LabeledAcProd} $^{225}\mathrm{Ac}$ production sample spectrum, processed with optimal filtering, with clear indication of $^{227}\mathrm{Ac}$  impurity visible from $^{223}\mathrm{Ra}$ , $^{227}\mathrm{Th}$ , and $^{211}\mathrm{Bi}$  daughters.}
\end{figure}

\begin{figure}[hbt]
\centering
\includegraphics[width=\linewidth]{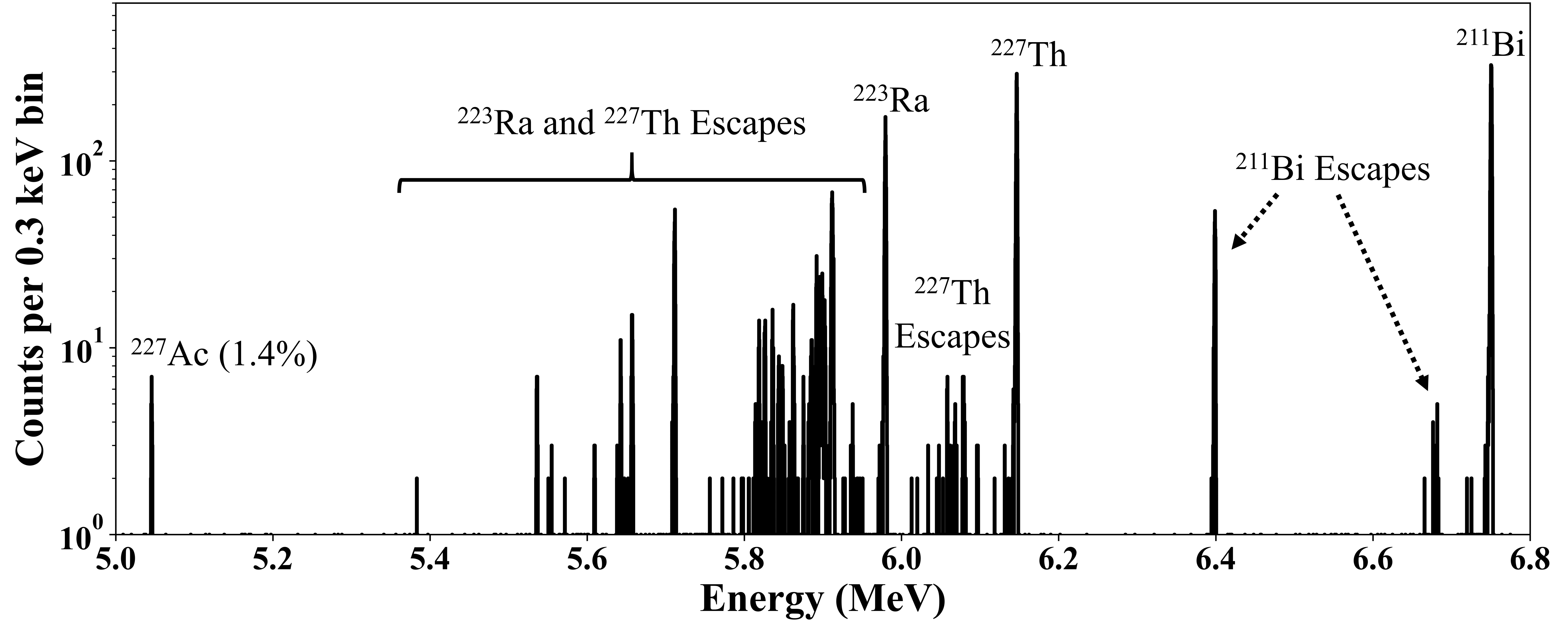}
\caption{\label{fig:227_spectrum} $^{227}\mathrm{Ac}$ spectrum, processed with optimal filtering, indicating the complicating presence of $\gamma$ and x-ray escapes as well as the high resolution achievable via DES. The $^{211}\mathrm{Bi}$ peak seen at 6.4 MeV has a full-width at half-maximum of 1.4 keV when fit with a single-tailed Bortels function.}
\end{figure}

 Sharp and easily-identified full-energy signature peaks are observed for many of the $\alpha$-emitting radionuclides, including $^{225}\mathrm{Ac}$, $^{221}\mathrm{Fr}$, $^{213}\mathrm{Bi}$, $^{227}\mathrm{Th}$, $^{223}\mathrm{Ra}$, and $^{211}\mathrm{Bi}$, but not $^{217}\mathrm{At}$, $^{213}\mathrm{Po}$, $^{219}\mathrm{Rn}$, or $^{215}\mathrm{Po}$. Several of the labeled peaks were used for calibration in the production sample and in the pure $^{227}\mathrm{Ac}$ sample, with plots of the filtered pulse height versus the corresponding energy of the peaks seen in Figure \ref{fig:Calibration}. The calibration was done using a cubic spline interpolation between the centroids of the peaks to account for the inherent non-linearity of the TES detectors, which resulted in a reasonably linear relationship between filtered pulse height and peak energy. For both $^{227}\mathrm{Ac}$ and $^{225}\mathrm{Ac}$ signals in the collected spectra, the slow decay time of the TES detectors led to pulse pileup that yielded distinct high-energy, poorly resolved features (Figure \ref{fig:Cascades}). Due to the abnormal shapes and energies, identification of these features was more challenging than the identification of the single direct decays, such as the labeled peaks seen in Figures \ref{fig:LabeledAcProd} and \ref{fig:227_spectrum}. However, identification was possible as we were able to extrapolate the linear energy calibration into the high-energy cascade region and couple this energy information with the half-life information seen in Figure \ref{DecayChains}.

\begin{figure}[hbt]
\centering
\includegraphics[width=\linewidth]{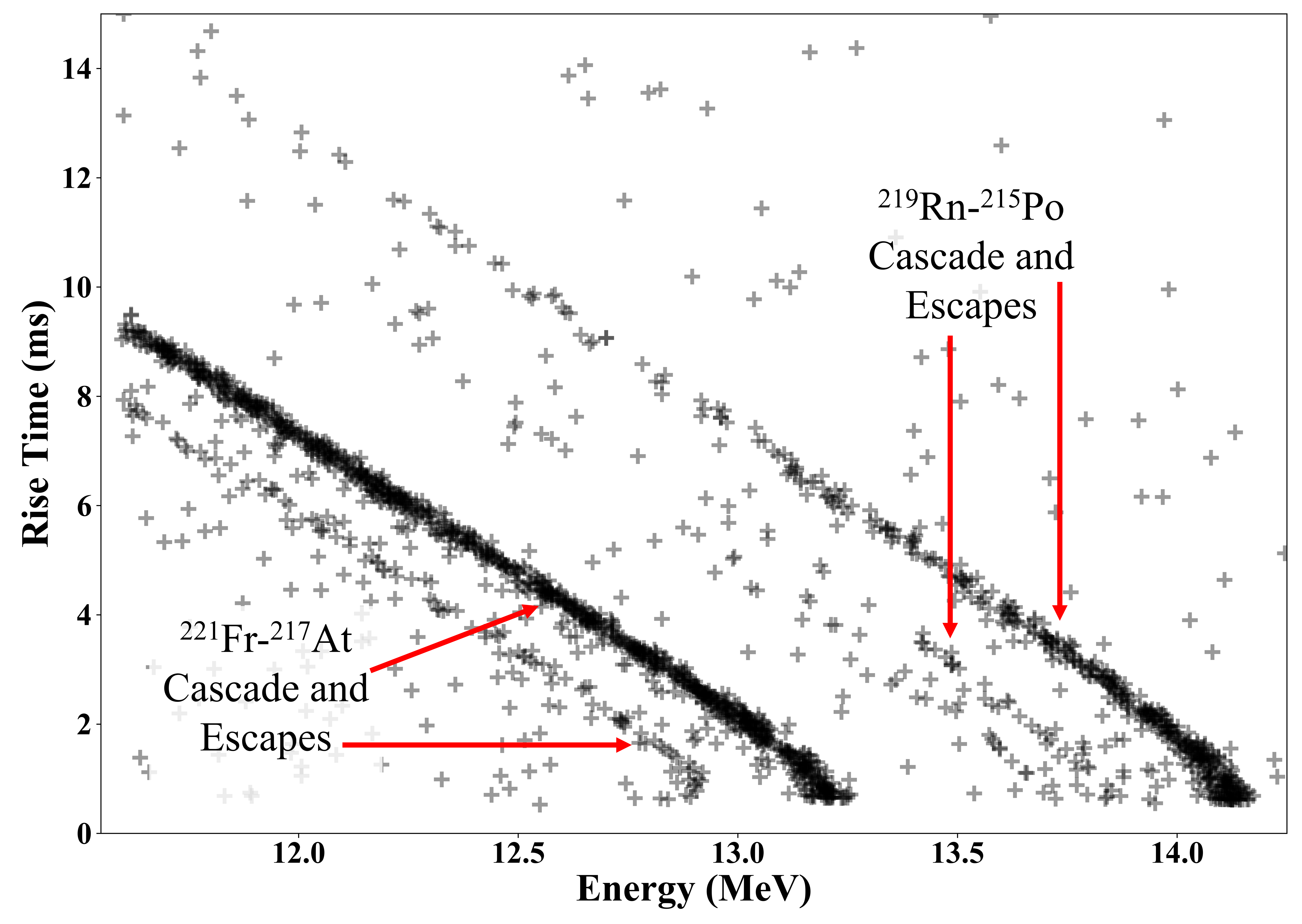}
\caption{\label{fig:Prod} Rise time versus pulse energy for the high energy cascade features seen in the $^{225}\mathrm{Ac}$ production sample. The $^{221}\mathrm{Fr}\rightarrow{}^{217}\mathrm{At}$ cascade is visible on the left, with an escape visible corresponding to a 218 keV escape from $^{221}\mathrm{Fr}$.  On the right is the $^{219}\mathrm{Rn}\rightarrow{}^{215}\mathrm{Po}$ cascade, with escapes from the $^{219}\mathrm{Rn}$ decay visible.}
\end{figure}

\begin{figure}[h!]
\centering
\includegraphics[width=\linewidth]{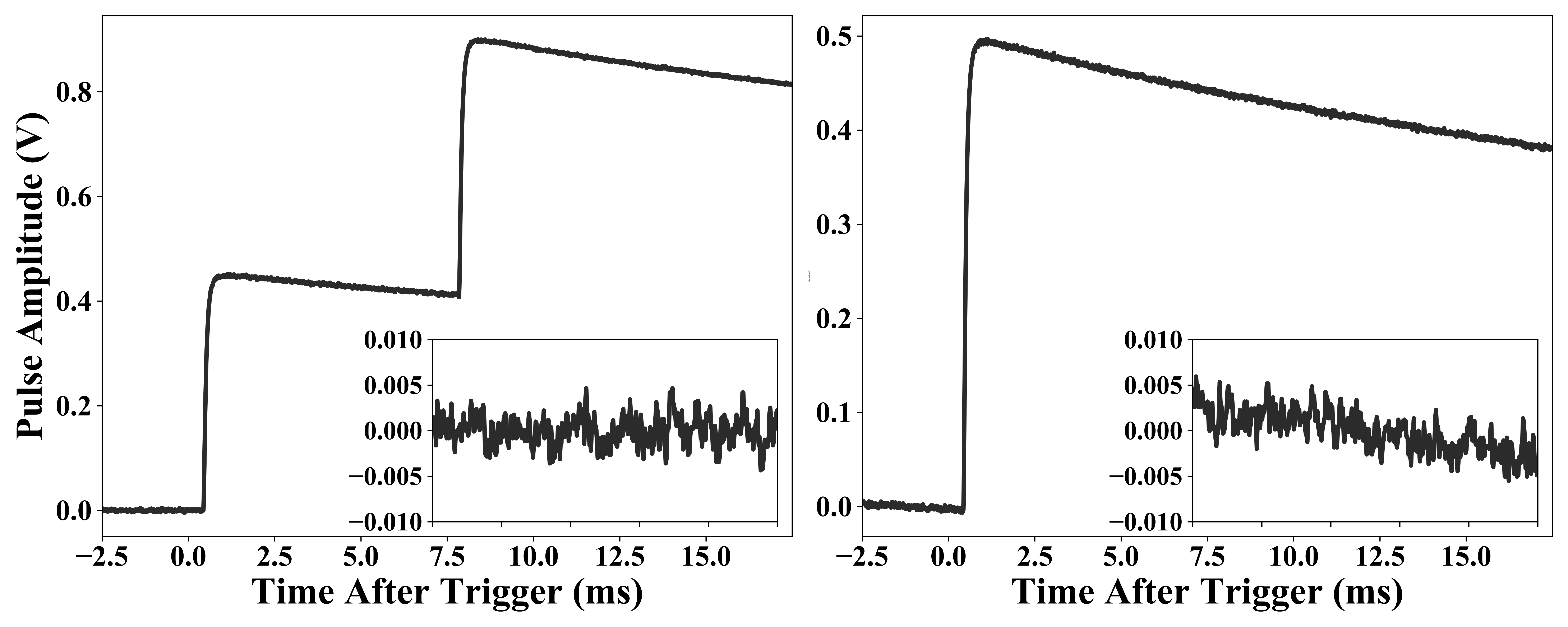}
\caption{\label{fig:AtPileup} Types of pileup visible in the $^{221}\mathrm{Fr}\rightarrow{}^{217}\mathrm{At}$ decay. (Left) Classic pileup seen in the pulse record of the $^{221}\mathrm{Fr}$ decay. (Left Inset) Pre-trigger baseline of the pulse shown in (Left). (Right) Pileup with a longer delay between the $^{221}\mathrm{Fr}$ and $^{217}\mathrm{At}$ than the post-trigger pulse record time. This resolves two distinct pulses, with the second pulse triggering during the decay time of the first pulse. (Right Inset) Sloped pre-trigger baseline for the pulse shown in (Right). This sloped baseline for the second event reduces achieved spectral resolution due to an improperly measured pulse height.}
\end{figure}

One high-energy cascade feature, seen in black in Figure \ref{fig:Cascades}, was visible in the pure $^{225}\mathrm{Ac}$ and mixed spectra but not the pure $^{227}\mathrm{Ac}$ spectrum, was identified as part of the $^{221}\mathrm{Fr}\rightarrow{}^{217}\mathrm{At}$ $\alpha$ decay cascade. When the rise time of the pulse records is plotted versus the peak pulse value, an escape feature at 11\%  of the intensity of the $^{217}\mathrm{At}$ decay is visible (Figure \ref{fig:Prod}).

The highest energy cascade feature, visible in the $^{227}\mathrm{Ac}$ and mixed spectra but not the pure $^{225}\mathrm{Ac}$ spectrum, was identified as the summed energy from the $^{219}\mathrm{Rn}$ and $^{215}\mathrm{Po}$ decays, which occur in rapid sequence. When pulse rejection is applied based on anomalously long rise times that correspond to visually clear pileup, three well-resolved peaks are observed within this feature (not shown). Measurable pileup signatures following rejection is possible due to the 1.8 ms half-life of $^{215}\mathrm{Po}$, yielding pileup that is not resolvable from normal pulses in rise-time space. These peaks have areas that correspond to the relative intensities of the full energy decay, the decay with a  7\% 402 keV $\gamma$-escape, and the decay with a 11\% 271 keV $\gamma$-escape from excited states of the $^{215}\mathrm{Po}$ daughter, with all other decay branches occurring with less than 1\% probability. This data rejection produces clean peaks but discards a large and unknown fraction of the signal, so is of little value to our quantification goal. But as we shall show, the uncut data from this region remain useful for quantification.

The $^{221}\mathrm{Fr}$ decay also appears as an isolated peak due to the fact that the $^{217}\mathrm{At}$ half-life is 32.3 ms, which is an order of magnitude shorter than the decay time of the detector and only 14.8 ms longer than the recorded sample time of each pulse. This leads to many well-resolved $^{221}\mathrm{Fr}$ pulses with $^{217}\mathrm{At}$  pulses arriving on the tail of this first pulse as well as pulses that are seen as obvious pileup, with the $^{217}\mathrm{At}$ pulse visible on top of the triggering pulse from the $^{221}\mathrm{Fr}$ decay (Figure \ref{fig:AtPileup}).  The second case yields a poorly resolved peak near the expected $^{217}\mathrm{At}$ energy, as the TES has not fully returned to thermal equilibrium from the $^{221}\mathrm{Fr}$ decay event by the time the $^{217}\mathrm{At}$ decay event has occurred. Because of these effects, neither the $^{221}\mathrm{Fr}$ nor the $^{217}\mathrm{At}$ decay events are of any value to our quantification goal.

\subsection{Quantification Method}

The measurements of the irradiated Th target sample occurred 47--61 days after EOB. Quantification was performed using a single peak from the $^{225}\mathrm{Ac}$  decay chain, and three separate signatures from the $^{227}\mathrm{Ac}$  decay chain. These peaks were selected due to the minimal interference between other peaks in the spectrum and, except for the case of $^{227}\mathrm{Th}$, their minimal escapes. For the direct Q-value peaks, region-of-interest (ROI) summing was used for quantification with a constant background estimated using 20 keV regions on either side of the peak. The net areas of the full-energy peaks were extracted and adjusted by a correction factor that is specific to each radionuclide. This correction factor accounts for photons from excited daughter states that may escape completely or deposit less than their full energy by stimulating additional Au x-ray emission, leading to a reduced area of the full-energy decay peak. The factor was determined using published branching ratios~\citep{ENSDF} for each decay scheme and modeled capture probabilities in the Au absorber by assuming a plane of gold 0.156 mm thick, corresponding to the thickness of the absorber. Decays may also produce conversion electrons which can generate secondary X-rays in the Au foil; these are usually reabsorbed but have a small possibility of escape.

From these measurements, the activities of $^{225}\mathrm{Ac}$ and $^{227}\mathrm{Ac}$ at the time of the measurements are deduced. The $^{227}\mathrm{Ac}$ daughters have not had time to reach equilibrium, so a factor is applied from solutions of the Bateman equations for the decay chains. We assumed that chemical purification was completed instantaneously 15 days after EOB. The EOB $^{227}\mathrm{Ac}$/$^{225}\mathrm{Ac}$ ratio was projected from these results.

The peak chosen for the quantification of $^{225}\mathrm{Ac}$ is the $^{225}\mathrm{Ac}$ direct $\alpha$-decay peak (Q = 5.9 MeV). This peak does not exhibit any interferences from the $^{227}\mathrm{Ac}$ decay chain, and no pileup occurs due to the 4.8 minute half-life of the $^{221}\mathrm{Fr}$ daughter. It does, however, have a complex decay scheme leading to a correction factor that is challenging to calculate; our model results suggest that between 2.7\% and 4.3\% of decays will not contribute to the full-energy peak. Its photon escape peaks are intermixed with those from $^{227}\mathrm{Th}$ and $^{223}\mathrm{Ra}$ and are not extractable. No other signature is practical for quantifying $^{225}\mathrm{Ac}$.

 The three signals selected for $^{227}\mathrm{Ac}$ quantification were the $^{211}\mathrm{Bi}$ (Q = 6.8 MeV) peak, the $^{227}\mathrm{Th}$ (Q = 6.5 MeV) peak, and the high energy cascade peak from the pileup of $^{219}\mathrm{Rn}$ (Q = 6.9 MeV) and $^{215}\mathrm{Po}$ (Q = 7.5 MeV, ${t}_{1/2}$ = 1.8 ms).
 
The $^{211}\mathrm{Bi}$ peak appears in a low background location in the spectrum and has one excited daughter state at 351.07 keV with a 13\% probability. The escape peak due to 351.07 keV photons would appear in the tail of the primary $^{221}\mathrm{Fr}$  peak from the $^{225}\mathrm{Ac}$ decay chain, and thus cannot be accurately measured. The modeled correction factor accounts for a 12.6\%--16.2\% loss of peak area. The full-energy peak lies on a very long low-energy tail from the $^{217}\mathrm{At}$ peak, a tail that exists due to slow detector decay time as mentioned previously. This background is easily quantifiable, but does limit this peak's usefulness for detection of very low $^{227}\mathrm{Ac}$ impurity.

The full-energy $^{227}\mathrm{Th}$ peak lies in an area with low background and no interference from other peaks. However, quantification is complicated by the high number of excited daughter states, with only 24.2\% of decays proceeding directly to ground state, requiring a substantial correction factor to be applied.  Our model suggests that 37.5\%--39.5\% of the peak area is lost to photon escape. Fortunately, this is corroborated by a measurement of the pure $^{227}\mathrm{Ac}$ sample, in which the escape peaks are easily measured (Figure \ref{fig:227_spectrum}), and also by comparison with the $^{211}\mathrm{Bi}$ peak, so we consider the correction factor to be reliable.

Quantification was also performed using the high energy $^{219}\mathrm{Rn}\rightarrow{}^{215}\mathrm{Po}$ $\alpha$-cascade sum for $^{227}\mathrm{Ac}$ activity. Due to overlap between the tail of this pileup feature and a partial pileup feature for the $^{221}\mathrm{Fr}\rightarrow{}^{217}\mathrm{At}$  cascade, the counts from each feature could not be resolved in one dimension (Figure \ref{fig:Cascades}). To combat this, the data from this overlap region was examined in a two-dimensional rise-time vs. peak height space. This produced distinct, well defined populations (Figure~\ref{fig:Prod}), allowing for the accurate quantification of $^{227}\mathrm{Ac}$  activity, plus lower-energy bands corresponding to photon escapes from two  excited states. Quantification was done by integrating a narrow parallelogram encompassing the primary structure, which is the full-energy $^{219}\mathrm{Rn}\rightarrow{}^{215}\mathrm{Po}$ decay, and lower energy cascade structures corresponding to $\gamma$-escapes from the $^{219}\mathrm{Rn}$ decay in the cascade process. Photon escape losses are negligible for $^{219}\mathrm{Rn}$, which decays to ground or one of the two excited states 99.8\% of the time, and also negligible for $^{215}\mathrm{Po}$, which decays to ground state 100\% of the time. The other pulses seen in scattered on either side of the main feature are pileup events from other decays, which contribute to a small background in this region. This background was estimated by averaging on either side of the  $^{219}\mathrm{Rn}\rightarrow{}^{215}\mathrm{Po}$ structure and was used to adjust the total $^{219}\mathrm{Rn}\rightarrow{}^{215}\mathrm{Po}$ counts measured.

\section{Results and Conclusion}

\begin{figure}[bth]
\centering
\includegraphics[width = \linewidth]{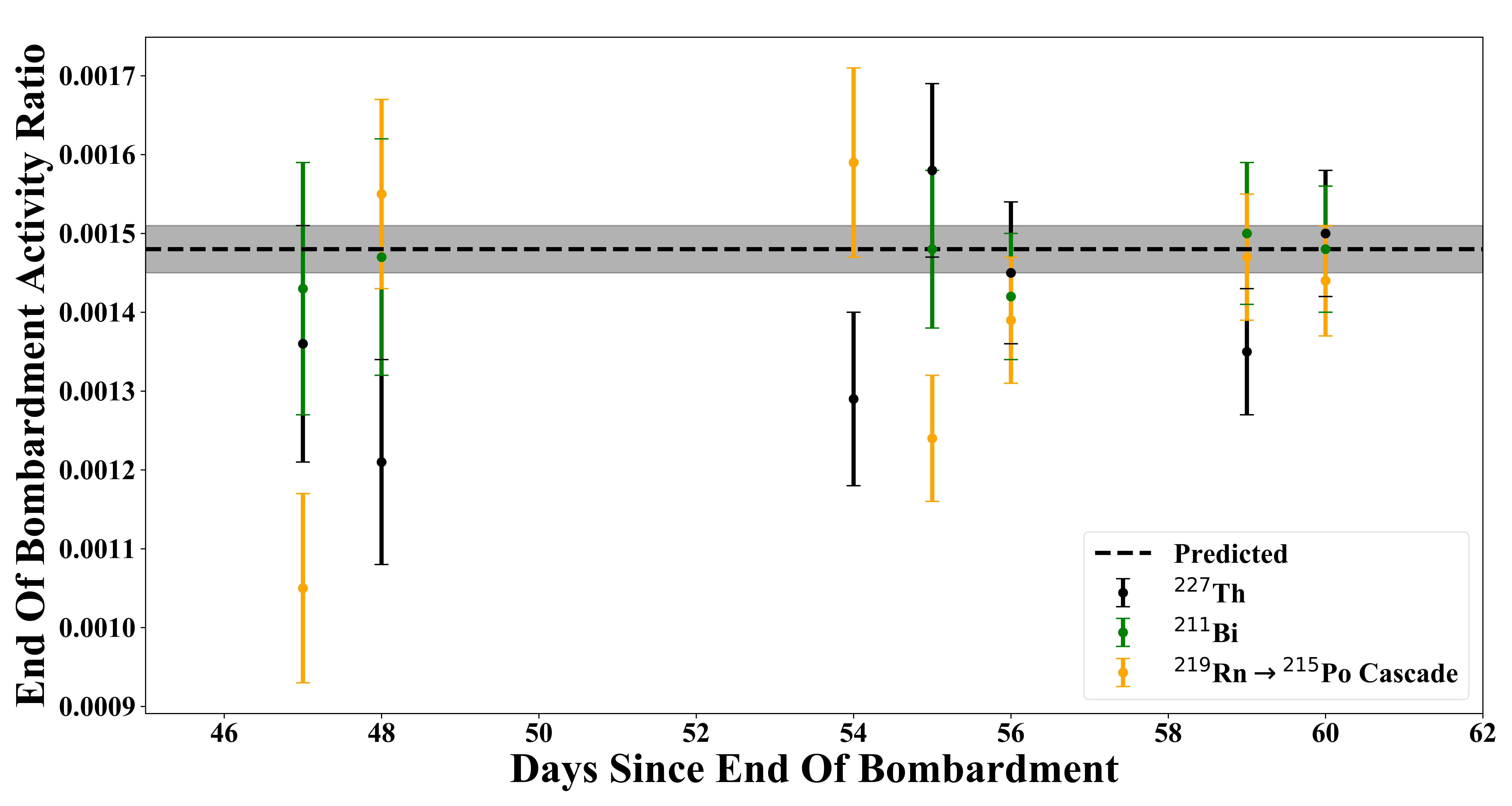}
\caption{\label{fig:FinalPercentagePlot} Plot indicating EOB $^{227}\mathrm{Ac}$/$^{225}\mathrm{Ac}$ activity ratio extrapolated from our measurement. The shaded region indicates the statistical uncertainty of the predicted value. $^{227}\mathrm{Ac}$ activity was calculated using three signals from its decay chain, and $^{225}\mathrm{Ac}$ activity was calculated using its direct decay. Peaks described in the legend were used for calculating the sample activity ratios. The error bars given for each measurement are statistical uncertainties.}
\end{figure}

Based on our measurements of the production sample, the EOB $^{227}\mathrm{Ac}$/$^{225}\mathrm{Ac}$ activity ratio is measured to be (0.141 $\pm$ 0.004)\% using $^{227}\mathrm{Th}$, (0.148 $\pm$ 0.004)\% using $^{211}\mathrm{Bi}$, and (0.139 $\pm$ 0.003)\% using the $^{219}\mathrm{Rn}$/$^{215}\mathrm{Po}$ cascade. Each of these is the weighted average from the seven independent measurements (shown in Figure~\ref{fig:FinalPercentagePlot}), and reported uncertainties are statistical only. Our final result for the EOB activity ratio is (0.142 $\pm$ 0.005)\%, combining the above and including systematic uncertainty from modeling and processing schedule details. The predicted ratio, determined using publicly available proton-induced cross sections \cite{IAEAdata1,IAEAdata2,Otuka2015} and the irradiation schedule, is (0.148 $\pm$ 0.003)\%, so our results are in agreement within 1$\sigma$.

Improved precision for the later measurements, visible in Figure~\ref{fig:FinalPercentagePlot}, is unsurprising because the $^{225}\mathrm{Ac}$ is decaying away with a 9.92 day half-life which decreases background, while at the same time the activities of the $^{227}\mathrm{Ac}$ progeny are increasing. A slight time-dependent trend in the measured ratio is also visible, with earlier results showing a smaller ratio.  This is suggestive of a systematic bias that we do not fully understand, but which may depend on count rate or on our understanding of the production sample processing schedule.

Our goal was to develop a technique capable of detecting an EOB $^{227}\mathrm{Ac}$/$^{225}\mathrm{Ac}$ activity ratio of 0.15\% as close as possible to the EOB.  $^{227}\mathrm{Ac}$ directly produces only a very tiny and virtually undetectable signature from its 1.4\% $\alpha$ branch (Figure ~\ref{fig:227_spectrum}), so the best option involves detection of its $^{227}\mathrm{Th}$ daughter, whose ingrowth is characterized by an 18.7 day half-life.  At five days after chemical separation, $^{227}\mathrm{Th}$ will have reached 17\% of its secular equilibrium value while $^{219}\mathrm{Rn}$, $^{215}\mathrm{Po}$, and $^{211}\mathrm{Bi}$, further down the decay chain, will have reached only 4\%, and so are less favorable candidates for early measurements.
 
We deduce from our data that the a priori detection limit is 0.0026 Bq of $^{227}\mathrm{Th}$ per Bq of $^{225}\mathrm{Ac}$, assuming a 24-hour measurement using a single DES channel and 1 Bq of sample. Assuming the realistic conditions of chemical purification 15 days post-irradiation followed by measurement five days later, this corresponds to an EOB limit of detection $^{227}\mathrm{Ac}$/$^{225}\mathrm{Ac}$ activity ratio of 0.38\%.  In order to meet our sensitivity goal we must engage seven of our eight DES channels in a simultaneous measurement, giving a detection limit of 0.14\%.  Substantial improvements may be possible with better understanding and reduction of the background, and with faster DES sensors to allow for higher activity samples.

This demonstration has shown that DES can distinguish closely related isotopic features for relative activity ratio quantification due to its extremely high resolution. This technique has 100\% efficiency, and is capable of measuring sub-becquerel activity ratios with accuracy as well as the capability to measure complex decay chains, with half lives shorter than the rise and decay times of the detector. This ability to manage complex nuclear decays while maintaining high resolution and accuracy using extremely small amounts of material indicates the applicability of DES as a radiometric technique.

\section{Acknowledgements}
Work presented in this article was supported by Technology Evaluation \& Demonstration funding from Los Alamos National Laboratory.  Los Alamos National Laboratory is operated by Triad National Security, LLC, for the National Nuclear Security Administration of U.S. Department of Energy (Contract No. 89233218CNA000001).  The production sample was graciously provided by the U.S. Department of Energy Isotope Program, managed by the Office of Science for Isotope R\&D and Production.

\appendix



\bibliography{bibliography.bib}

\end{document}